# Tip-gating Effect in Scanning Impedance Microscopy of Nanoelectronic Devices

Sergei V. Kalinin[1] and Dawn A. Bonnell

Department of Materials Science and Engineering,

University of Pennsylvania, 3231 Walnut St, Philadelphia, PA 19104

Marcus Freitag and A.T. Johnson

Department of Physics and Astronomy and Laboratory for Research on the Structure of

Matter, University of Pennsylvania, 209 South 33rd St, Philadelphia, PA 19104

**ABSTRACT**

Electronic transport in semiconducting single-wall carbon nanotubes is studied by combined scanning gate microscopy and scanning impedance microscopy (SIM). Depending on the probe potential, SIM can be performed in both invasive and non-invasive mode. High-resolution imaging of the defects is achieved when the probe acts as a local gate and simultaneously an electrostatic probe of local potential. A class of weak defects becomes observable even if they are located in the vicinity of strong defects. The imaging mechanism of tip-gating scanning impedance microscopy is discussed.

---

[1] Currently at the Oak Ridge National Laboratory

Development and implementation of nano- and molecular electronic devices necessitates reliable techniques for device characterization. Current based transport measurements require contacts and generally do not allow spatial resolution. Therefore, significant attention has been focused recently on hybrid transport measurements by Scanning Probe Microscopy (SPM) based techniques.[1,2,3] SPM tips act as non-invasive moving dc (Scanning Surface Potential Microscopy) or ac (Scanning Impedance Microscopy) voltage probes similar to 4 probe resistance measurements.[4,5] Alternatively, in Scanning Gate Microscopy (SGM) tip induced changes in the circuit resistance are measured.[6,7] The resolution in SGM is limited and only strong defects can be located.[8] Here, we present a novel scanning impedance mode where the tip induces a local perturbation and, at the same time, acts as an electrostatic probe for the local potential. This tip-gating SIM enables us to detect weak defects in the nanodevices that are not resolved in SGM. The bias dependence of the contrast provides the information on the local electronic structure of the defects.

We illustrate this approach with a carbon nanotube circuit, in which we characterize defect mediated transport properties. $p$-doped carbon nanotubes are grown by catalytic chemical vapor deposition (CVD) on a $SiO_2$/Si substrate.[9] In SIM, the tip is held at constant voltage $V_{tip}$ and a lateral bias, $V_{lat} = V_{dc} + V_{ac}\cos(\omega t)$, is applied across the sample. The lateral bias induces an oscillation in the surface potential, $V_{surf} = V_s + V_{ac}(x)\cos(\omega t + \varphi(x))$, where $V_{ac}(x)$ and $\varphi(x)$ are the position dependent voltage oscillation amplitude and phase shift and $V_s$ is the dc surface potential. The amplitude and phase of the mechanical tip oscillations is used to obtain the amplitude and phase of local voltage oscillations, providing quantitative information on the local transport properties.[5] Simultaneous current measurements (SGM) allow the tip effect on transport through the nanotube to be estimated.

The tip-voltage dependence of the SGM images is shown in Figure 1 and a number of defects exhibit clear contrast variations. The apparent size of the defects increases linearly with tip voltage and the diameter of the spot in the SGM image is $D_i = 0.54\,RV_{tip}/V_i^*$, where $R$ is tip radius of curvature and $V_i^*$ is the depletion surface potential which corresponds to the onset of depletion at the $i$-th defect.[8] The depletion surface potential is related to the local band structure, providing much needed physical information on nanotube FET. Depending on the separation between the Fermi level and the top of the valence band in the semiconducting nanotube the defects can be classified as strong (small $V_i^*$) and weak (large $V_i^*$). The weak defects affect transport at higher tip bias, but the contrast from strong defects simultaneously increases overshadowing weak neighbors. Therefore, only strong defects are observed in SGM.

The local potential in the nanotube is provided by SIM. For a weak negative tip bias, SIM acts as a relatively non-invasive probe of nanotube conductance. Fig. 1(f) illustrates the potential distribution along a segment of the nanotube FET in a conductive (ON) state measured in a non-invasive way with $V_{tip}$ = -2 V. The potential drop in the device is uniform indicating that none of the defects is in the high-resistance state and the current is diffusive along the entire length of the nanotube. Current enhancement in the vicinity of the electrodes owing to the effect of Schottky barriers [Fig. 1a] is observed.[10,11] Using a positively biased back-gate depletes the defects and potential steps at the position of the defects are resolved.[8] However, the potential drop at the individual defect is necessarily a fraction of lateral bias applied across the system. From a simple voltage divider model, the potential drop at each defect is proportional to the defect resistance, resulting in small drops at weak defects. Moreover, if the defect spacing is small compared to the resolution of the SPM (~50-100 nm), the potential drops at individual defects cannot be resolved. A consequence is that transport through a line of closely spaced defects might be erroneously interpreted as diffusive.

In the case of SIM with a positive tip bias, tip gating significantly affects the potential distribution, resulting in a non-monotonic voltage profile across the nanotube as illustrated in Fig. 1(h-k). The mechanism of tip-gating SIM is illustrated in Fig. 2. When the tip is in location 1, defect 1 is in the OFF-state. The resistance of the segment of the nanotube to the left of the tip position, $R_l$, is large, whereas the resistance of the nanotube segment to the right of the tip position, $R_r$, is small. Therefore, the nanotube acts as a voltage divider and the local voltage oscillation amplitude $V(x) = V_{ac} R_r / (R_r + R_l)$ at position 1, is close to zero. If the tip is moved to position 2 the resistance of defect 1 is unaltered. However, now the tip is positioned above the biased region (high resistance region is to the right), and the oscillation amplitude is high. The width of the transition region is determined by the tip-surface transfer function and provides a direct measure of SIM resolution.[12] On moving to the left, the gating efficiency for defect 1 decreases and for defect 2 increases; therefore, the ratio $R_r / (R_r + R_l)$ also decreases resulting in a saw tooth-like signal. Specifically, in position 3 for the defects of equal strength, $R_r = R_l$ and therefore $V(x) = V_{ac}/2$ independent of defect conductance, while the current through the nanotube can be below the detection limit. The exact shape of the SIM profile in the tip-gating regime is determined by the convolution of the tip-induced surface potential distribution, the bias dependence of defect conductance, and the tip-tube capacitance and, therefore, represents a complex problem. However, the immediate implication is that even weak defects are directly observable in the SIM image and the associated depletion surface potentials can be determined.

These considerations can be further supported by a simple model. Assume that the resistance [Ohm/nm] of the non-defect part of the nanotube is $\alpha$ and the resistance of the defect is related to the turn-on voltage as $R_i = \alpha \exp[(V(x,y) + V_i^*)/kT]$, $kT = 0.027$ eV and $V(x,y)$ is tip-induced surface potential. Fig. 3 illustrates the potential distribution in the 1 μm nanotube with four defects when globally gated by the back electrode. The tip is assumed non-invasive and tip-broadening effect is accounted for by the convolution with a Lorentzian function with Full Width at Half Maximum (FWHM) of 36 nm. In this case, the potential steps corresponding to weak defects (Defect 2 and 3) are small (Fig. 3a). For large back gate biases, the potential distribution along the nanotube is relatively bias independent. Figure 3b shows the potential distribution in the nanotube when the local tip gate is used. While for small tip bias the tip effect on potential distribution in the nanotube is minimal, the large tip biases are associated with the onset of the tip-gating regime. In this case, each defect is associated with a maximum in voltage oscillation amplitude and can be clearly distinguished on SIM image. In principle, all defects can be distinguished in the SGM profile as well; however, the resistance of the nanotube in the off state is typically > 1 GΩ. In conjunction with the small dwell time (3-10 ms depending on scan speed) of the tip at any selected point this renders reliable SGM measurements in the low-conductance state and identification of weak defects extremely difficult. In contrast, the SIM image in the tip-gating regime is formed by a number of sharp spikes on a zero background (comp. Fig. 1 e,k). The resolution in SIM is proportional to the square of the tip surface transfer function and, therefore, is higher than that of the SGM.

In conclusion, we have compared the contrast of individual defects in semiconducting SWNTs in ac-scanning gate microscopy and scanning impedance microscopy. In the SIM tip-gating mode, the tip acts simultaneously as a gate and a voltage sensor and is sensitive to weak defects unobservable by SGM. The voltage threshold for contrast is a direct measure of the electronic structure of the defect. This approach is general and will be applicable in all cases of molecular- and nanocircuits in which an atomic defects and proximity interactions give rise to variations in local resistance.

We acknowledge the support from MRSEC grant NSF DMR 00-79909. The authors are grateful to Dr. M. Radosavljević (IBM) and Prof. M. Cohen (UPenn) for valuable discussions.

**FIGURE CAPTIONS**

**Figure 1**. SGM (a-e) and SIM (f-k) images of a SWNT at –2 V (a,f), 2 V (b,g), 4 V (c,h), 6 V (d,i) and 7 V (e,h) tip bias. DC bias dependence of SGM contrast allows defect strength to be quantified. In SIM, high tip bias results in "tip-gating" effect.

**Figure 2.** Schematic diagram of the tip-gating effect. D1 and D2 denote strong defects, D3 denote a weak defect. Locations 1-4 correspond to tip positions as discussed in the text. Shown are schematic SIM profiles for small (solid line, only D1 and D2 are in the OFF state) and large (dotted line, all defects are in the OFF state) tip bias.

**Figure 3.** Simulated potential profiles along the nanotube for a non-invasive probe for different back gate bias (a) and in the tip-gating regime for different tip gate bias (b). The depletion surface potentials for defects 1-4 are 100 mV, 50 mV, 20 mV and 100 mV, respectively. The profiles are calculated for tip-or back-gate induced potentials of 30 mV (solid), 100 mV (dashed), 300 mV (dotted), 1 V (dash-dot) and 3 V (dash-dot-dot).

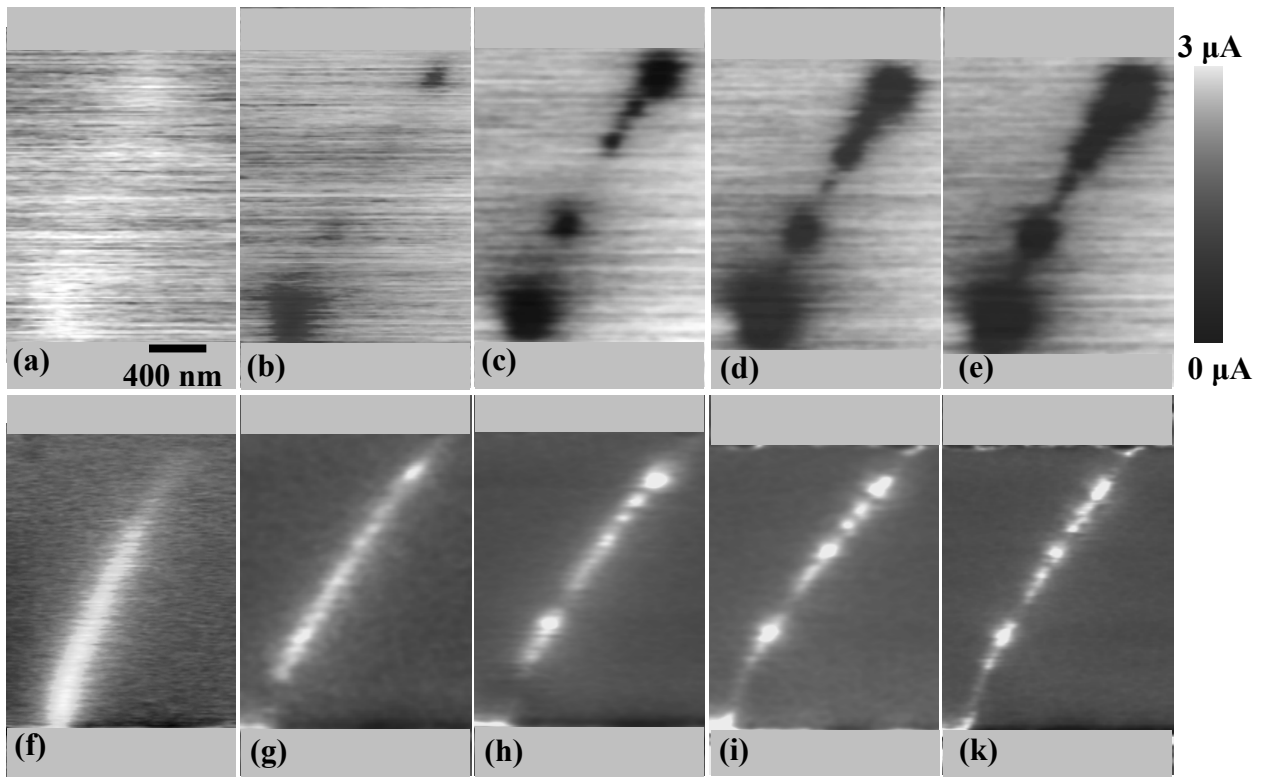

**Fig. 1.** S.V. Kalinin, D.A. Bonnell, M. Freitag, and A.T. Johnson

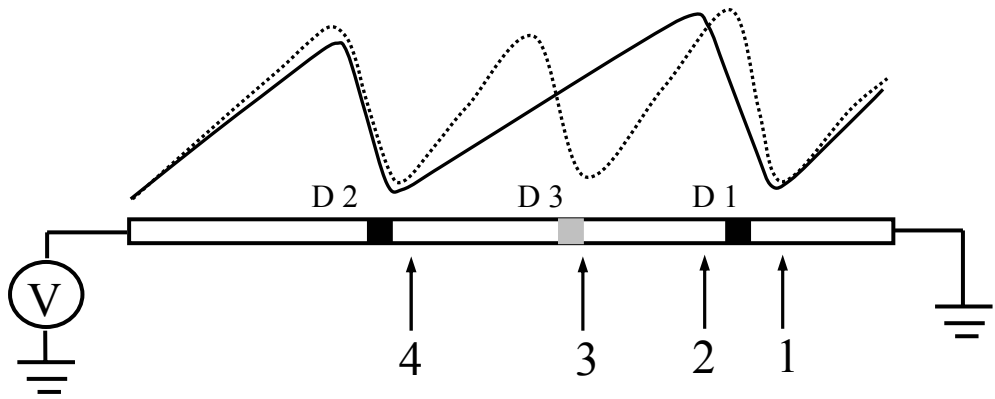

**Fig. 2.** S.V. Kalinin, D.A. Bonnell, M. Freitag, and A.T. Johnson

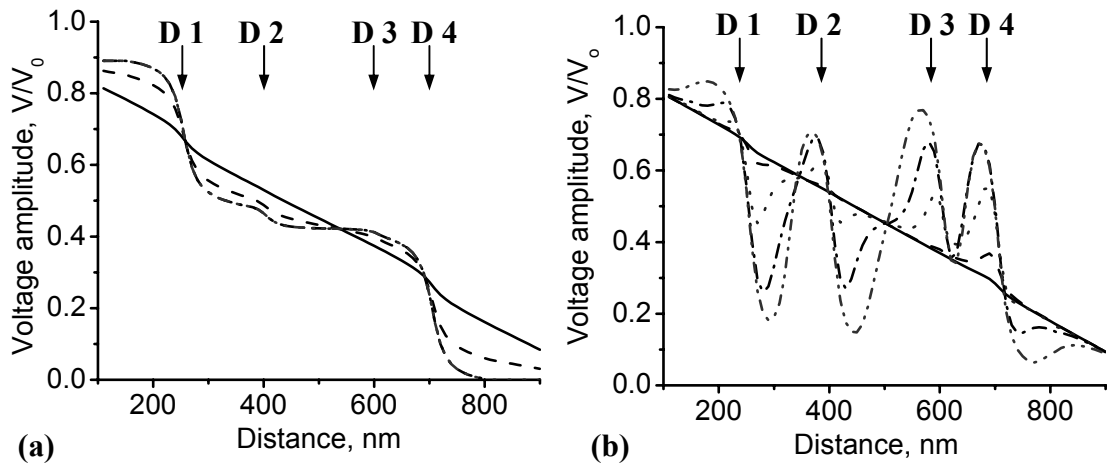

**Fig. 3.** S.V. Kalinin, D.A. Bonnell, M. Freitag, and A.T. Johnson